\documentclass[twocolumn,aps,prb,superscriptaddress]{revtex4-1}

\usepackage{amsmath}
\usepackage{graphicx}
\usepackage{color}
\usepackage{url}
\usepackage[left=0.75in, right=0.75in,top=0.75in,bottom=1.5in]{geometry}

\begin{document}

\title{Concentration tuned tetragonal strain in alloys: application to magnetic anisotropy of FeNi$_{1-x}$Co$_x$}

\author{Aleksander L. Wysocki}
\email{alexwysocki2@gmail.com}
\affiliation{Ames Laboratory, Ames, IA 50011, USA}

\author{Manh Cuong Nguyen}
\affiliation{Ames Laboratory, Ames, IA 50011, USA}

\author{Cai-Zhuang Wang}
\affiliation{Ames Laboratory, Ames, IA 50011, USA}

\author{Kai-Ming Ho}
\affiliation{Ames Laboratory, Ames, IA 50011, USA}

\author{Andrey V. Postnikov}
\affiliation{Universit\'e de Lorraine, Metz, F-57078, France}

\author{Vladimir P. Antropov}
\affiliation{Ames Laboratory, Ames, IA 50011, USA}

\date{\today}

\begin{abstract}
We explore an opportunity to induce and control tetragonal distortion in materials. The idea involves formation of a binary alloy from parent compounds having body-centered and face-centered symmetries. The concept is illustrated in the case of FeNi$_{1-x}$Co$_x$ magnetic alloy formed by substitutional doping of the L1$_0$ FeNi magnet with Co. Using electronic structure calculations we demonstrate that the tetragonal strain in this system can be controlled by concentration and it reaches maximum for $x=0.5$. This finding is then applied to create a large magnetocrystalline anisotropy (MAE) in FeNi$_{1-x}$Co$_x$ system by considering an interplay of the tetragonal distortion with electronic concentration and chemical anisotropy. In particular, we identify a new ordered FeNi$_{0.5}$Co$_{0.5}$ system with MAE larger by a factor 4.5 from the  L1$_0$ FeNi magnet.
\end{abstract}

\maketitle

Controlling tetragonal distortion in solids is a common approach to improve properties of functional materials. The tetragonal strain decreases the symmetry of the system allowing for a number of effects which are, otherwise, forbidden in cubic structures. From a different perspective, such distortions can be used to tune materials properties for desired applications or to induce a phase transition in the system creating a new functional phase.

Experimentally, tetragonal distortions are typically realized by coherent growth of the material on a lattice-mismatched substrate or buffer. However, such distortions can exist only for ultrathin layers since for thicker films the strain is released by formation of dislocations. Therefore, this approach is unsuitable when large samples are required for applications. For bulk systems one can attempt to create tetragonal strain by interstitial doping with small atoms (i.e., C or B) but such approach is difficult to control and usually only modest strains can be achieved.

The problem of controlling tetragonal strains in cubic systems is of great importance in the field of permanent magnetism. Due to recent supply shortage of rare-earth elements it is, currently, crucial to design new rare-earth-free and high-energy-product permanent magnets.\cite{Lewis,McCallum} From this perspective, transition metal magnets (Fe, Co, Ni, and their alloys) are especially promising class of materials since their relatively high magnetization could potentially lead to large energy products. In addition, these materials typically have large Curie temperatures that makes them ideal for high temperature operations. Unfortunately, transition metal magnets often crystallize in cubic structures for which the second order contribution to the magnetocrystalline anisotropy energy (MAE) is zero by symmetry. This results in a low MAE ($\sim1$ $\mu$eV/atom) which severely limits applications of these materials as permanent magnets. A notable exception is a marginally stable form of FeNi called tetrataenite\cite{Clarke} that has a tetragonal L1$_0$ structure and a sizable MAE ($\approx$ 40 $\mu$eV/atom).\cite{Shima} This suggests that a much larger MAE could be realizes in unstable families of transition metal magnets with a built-in tetragonal distortion. This concept was supported by electronic structure calculations.\cite{Burkert,Burkert2} In particular, MAE as large as 800 $\mu$eV/atom has been predicted for a strained ($c/a\approx 1.23$) FeCo system.\cite{Burkert2} Large MAE was, indeed, observed for ultrathin layer of strained FeCo epitaxially grown on lattice-mismathed substrate.\cite{Andersson,Yildiz,Yildiz2}

Here, we propose a strategy for tuning tetragonality in materials by mixing compounds with body- and face-centered symmetries. As a realization of this idea, we consider formation of a FeNi$_{1-x}$Co$_x$ alloy from B2 FeCo and unstable L1$_0$ FeNi parent compounds. Using first principles electronic structure calculations we demonstrate that the tetragonal strain can be naturally tuned by concentration with a maximum around $x=0.5$. Further, the MAE FeNi$_{1-x}$Co$_x$ is analyzed. We show that for a random alloy MAE remains low despite the presence of strong tetragonal distortion. However, a large MAE (180 $\mu$eV/atom) can be achieved for an ordered structure created by vertical stacking of FeNi and FeCo layers.

\begin{figure}[b!]
\includegraphics[width=1.4\hsize]{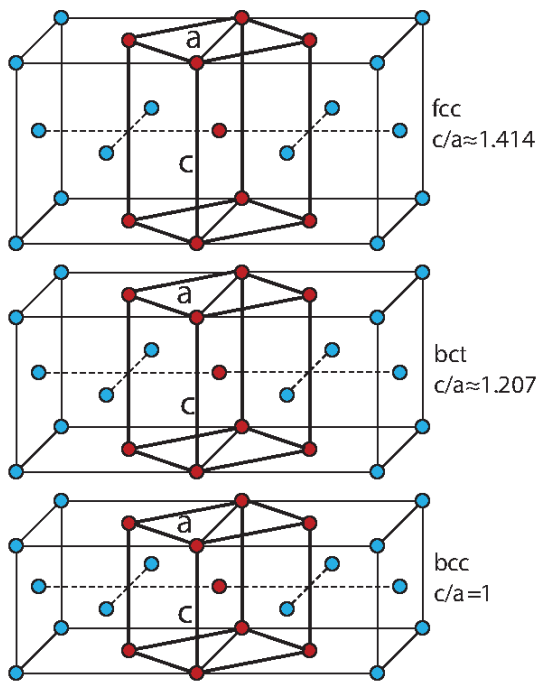}
\caption{Relation between fcc and bcc lattices. (top) fcc is equivalent to the bct lattice with $c/a=\sqrt{2}$. Compression of the lattice until $c/a=1$ (Bain transformation) results in cubic bcc lattice (bottom). For intermediate values of $c/a$ (middle) the lattice has tetragonal symmetry.}
\label{Fig1}
\end{figure}

The key idea follows from the observation that the face-centered cubic (fcc) lattice can be viewed as a body centered tetragonal (bct) lattice with the $c/a$ ratio equal to $\sqrt{2}$ (see Fig.~\ref{Fig1}). Consequently, the body centered cubic (bcc) lattice can be obtained from the fcc lattice by the appropriate compression so that $c/a=1$. Both for $c/a=1$ and $c/a=\sqrt{2}$ the system has a cubic symmetry. However, for intermediate $c/a$ values the solid has a tetragonal distortion. This transformation is known as continuous Bain transformation and has been observed experimentally. \cite{Bain} Let us now consider two cubic (or nearly cubic) material with similar atomic volumes but different structures. One has a face-centered symmetry and the other one has a body-centered symmetry. According to the above discussion, we can expect that an alloy formed from these two materials has a crystal structure corresponding to an intermediate point along the Bain transformation path with a tetragonal strain that is controlled by concentration. 

The above concept can be realized in the FeNi$_{1-x}$Co$_x$ alloy formed by substitutional doping of the L1$_0$ FeNi magnet with Co at the Ni site. Within the bct lattice L1$_0$ FeNi has $c/a \approx 1.424$ that is slightly larger from the ideal fcc value ($c/a=\sqrt{2}$).\cite{Kotsugi} If all Ni atoms are replaced by Co, we obtain FeCo intermetallic compound with the CsCl structure. This system has a cubic body-centered symmetry with $c/a=1$. According to the discussion in the previous paragraph, we, therefore, expect that for partial Ni-Co substitutions the resulting FeNi$_{1-x}$Co$_x$ alloy will develop a sizable tetragonal distortion. Below, we investigate this system using first principles electronic structure calculations.

\begin{figure}[t!]
\includegraphics[width=1\hsize]{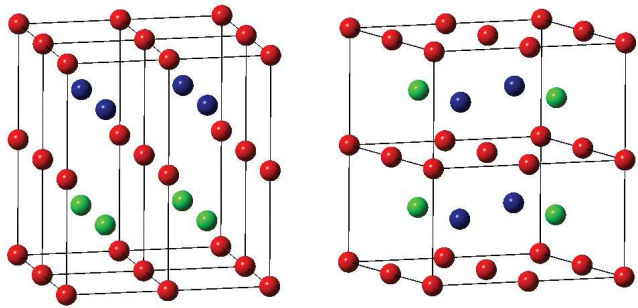}
\caption{Schematic illustration of supercell models for disordered (right) and ordered (left) Ni$_{1-x}$Co$_x$ alloy in the case of $x=0.5$. The red, green, and blue spheres denote Fe, Ni and Co atoms, respectively.}
\label{Fig2}
\end{figure}

In order to model doping we used 2$\times$2$\times$1 and 1$\times$1$\times N$ ($N=2,3,4$) supercells with respect to the primitive bct cell of the parent compounds. The 2$\times$2$\times$1 supercell was used to simulate the random alloy (Co atoms randomly substitute Ni atoms), see Fig. \ref{Fig2} (right) in the case of $x=0.5$. In addition, we also considered the case of ordered Ni$_{1-x}$Co$_x$ alloy in which some of Ni layers in L1$_0$ FeNi are replaced  by Co. As a result, the structure consists of vertical stacking of L1$_0$ FeNi and B2 FeCo layers, see Fig. \ref{Fig2} (left).

The calculations were performed using the density functional theory with PBE exchange-correlation functional. The Kohn-Sham equations were solved using the projector augmented wave method \cite{Blochl} as implemented in the VASP code\cite{Kresse,Kresse2}.  The cutoff energies for the plane wave and augmentation charge were 270 eV and 545 eV, respectively. For the primitive bct cell we used 12$\times$12$\times$12 24$\times$24$\times$24 $\Gamma$-centered k-point mesh for relaxation and anisotropy calculations, respectively. For larger cells the k-point mesh was scaled accordingly. The lattice parameters and the ionic positions have been relaxed until the Hellmann-Feynman forces were converged to less than 0.01 eV/\AA. MAE has been calculated using the force theorem. The site-resolved MAE were calculated using the approach described in Ref.~\onlinecite{Antropov}.

\begin{figure}[t!]
\includegraphics[width=1\hsize]{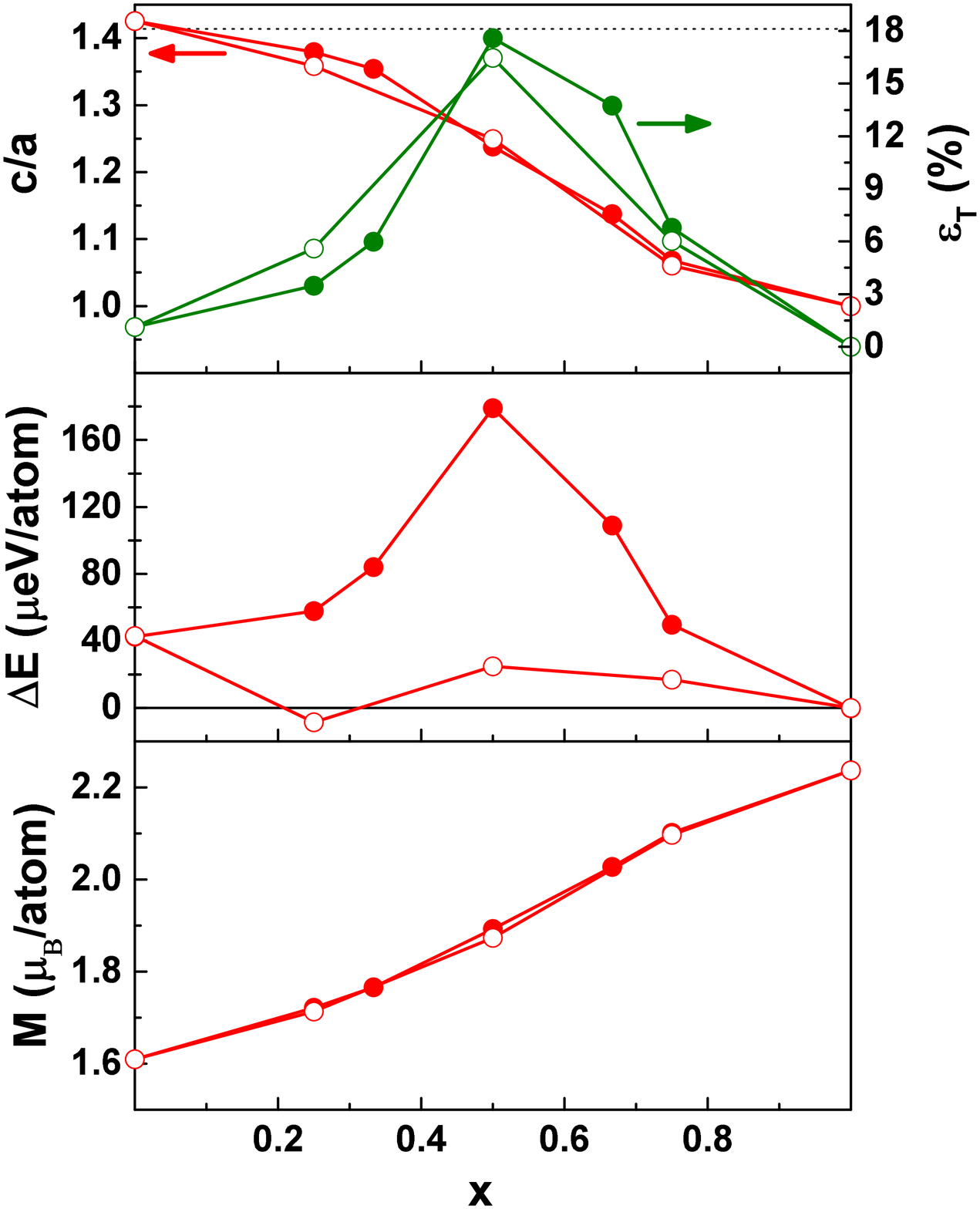}
\caption{Calculated $c/a$ ratio, tetragonal distortion parameter (top), MA energy (middle), and spontaneous magnetization (bottom) of FeNi$_{1-x}$Co$_x$ as a function of the Co content. Open (filled) circles correspond to the random (ordered) alloy case. The dotted horizontal line denotes $c/a=\sqrt{2}$ that corresponds to the face-centered cubic symmetry.}
\label{Fig3}
\end{figure}

Figure \ref{Fig3} (top) shows the concentration dependence of $c/a$ ration and the tetragonal strain both for disoredered and ordered FeNi$_{1-x}$Co$_x$ alloy. Here, the tetragonal strain is defined as

\begin{equation}
\epsilon_T=\min\left(|c/a-\sqrt{2}|,|c/a-1|\right)\times100\%
\end{equation}

We observe that under doping the $c/a$ smoothly decreases from nearly fcc value down to the bcc value. Consequently, the tetragonal distortion develops for intermediate dopings. In particular, $\epsilon_T$ depends strongly on concentration and it reaches maximum of 18\% at $x=0.5$. Importantly, we find that both $c/a$ and $\epsilon_T$ have a weak dependence on doping configuration. In fact, both disordered and ordered FeNi$_{1-x}$Co$_x$ alloy have similar a concentration dependence of $c/a$ and $\epsilon_T$. These results indicate that the tetragonal distortion in FeNi$_{1-x}$Co$_x$ alloy can be controlled by doping concentration.

Let us now consider the magnetic properties of FeNi$_{1-x}$Co$_x$ alloy. The calculated spontaneous magnetization as a function of concentration is shown in Fig. \ref{Fig3} (bottom). As seen, the magnetization is virtually independent on doping configuration and it increases smoothly with $x$ starting from 1.6 $\mu_B/$atom value for FeNi to 2.3 $\mu_B/$atom for FeCo.

\begin{figure}[t!]
\includegraphics[width=1\hsize]{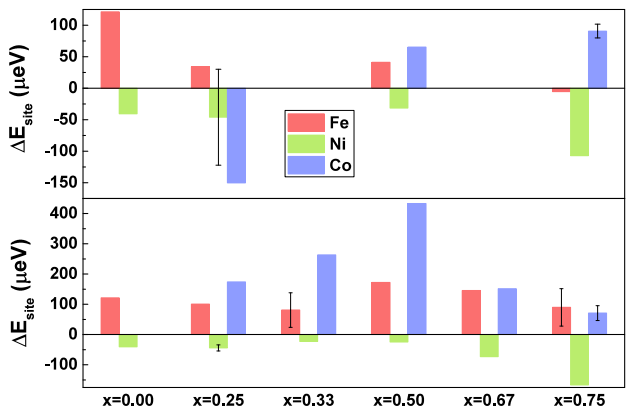}
\caption{Concentration dependence of the site-resolved MAE for disordered (top) and ordered (bottom) FeNi$_{1-x}$Co$_x$ alloy. The shown values are averaged over different nonequivalent atoms of the same specie in the considered supercells. The error bars denote the corresponding standard deviation when the latter is significant.}
\label{Fig4}
\end{figure}

In the case of MAE the situation is much more complicated. Concentration dependence of MAE is shown in the middle panel of Fig. \ref{Fig3} both for disordered and ordered alloy. As discussed in the introduction, the development of a strong tetragonal strain for intermediate doping concentrations may lead to an enhancement of MAE. For disordered FeNi$_{1-x}$Co$_x$ alloy, however, this is not the case. In fact, in this case doping with Co results in MAE being even smaller than for L1$_0$ FeNi compound despite the increase of $\epsilon_T$. This result reflects a fragile nature of MAE which, in addition to tetragonal strain, depends also strongly on electronic concentration and chemical anisotropy in the system. The complex interplay between these three factors for FeNi$_{1-x}$Co$_x$ alloy can be illustrated by plotting site-resolved MAE \cite{Antropov,FEN} as a function of Co concentration, see Fig.~\ref{Fig4}. As seen, for pure FeNi, the Fe atom has a large contribution to MAE above 100 $\mu$eV. On the other hand, the increase of electronic concentration for the Ni atom leads to a significant negative contribution to MAE resulting in a rather moderate total MAE for this compound. In the case of the disordered FeNi$_{1-x}$Co$_x$ alloy, the Co doping results in a strong reduction of the Fe contribution which becomes even negative for large $x$. On the other hand, the Ni contribution to MAE changes somehow with concentration but it always remains negative. The Co atoms contribution to MAE varies strongly with $x$ starting from large negative values for low concentrations to moderate positive values for larger $x$. Note that the electronic concentration for each specie remains approximately the same as $x$ increase. Therefore, since the tetragonal distortion remains positive for all concentrations, we can conclude that this is the chemical anisotropy mechanism that is responsible for the reduction of the Fe contribution and the strong variation of the Co contribution with doping. Indeed, for the disordered alloy the atomic environment of Fe and Co atoms changes with doping. The chemical anisotropy mechanism can be, thus, controlled by alloy ordering. In the case of the layered ordering shown in Fig. \ref{Fig2} (left) the chemical anisotropy of the parent compounds is preserved. The corresponding site-resolved contributions to MAE are shown in Fig. \ref{Fig4} (bottom). As seen, the Ni contribution as a function changes somewhat as a function of Co concentration but, similarly as in the case of disordered alloy, it remains negative for all dopings. This indicates that the Ni contribution is primarily controlled by the electronic concentration. However, the concentration dependence of both Fe and Co contribution changes completely when the atomic ordering is introduced. Indeed, the Fe contribution has a rather weak doping dependence and remains large and positive for all $x$. Therefore, the chemical anisotropy mechanism is crucial in this case. The Co contribution is also positive for all $x$ but its magnitude changes significantly with the concentration. and the dependence roughly follows $\epsilon_T$. More specifically, the concentration dependence for large and small $x$ the Co contribution is close to the Fe one but for intermediate doping concentrations the Co contribution shows a strong increase and for $x=0.5$ it reaches a gigantic value above 400 $\mu$eV. This indicates that the Co contribution is controlled both by both tetragonal distortion and chemical anisotropy mechanisms.

\begin{figure}[t!]
\includegraphics[width=1\hsize]{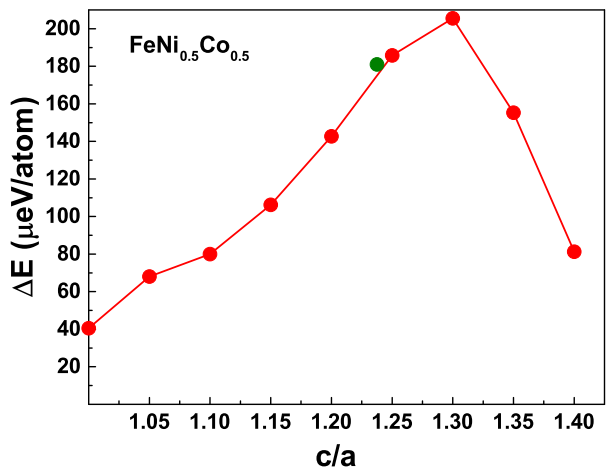}
\caption{MAE of the ordered FeNi$_{0.5}$Co$_{0.5}$ alloy as a function of the $c/a$ ratio. The green circle denotes MAE for equilibrium $c/a$.}
\label{Fig5}
\end{figure}

As a result of the strong increase of the Fe and Co contributions, the total MAE of the ordered FeNi$_{1-x}$Co$_x$ alloy is significantly larger than in the case of the disordered alloy (see Fig. \ref{Fig3}). More importantly, for the ordered alloy the MAE dependence on concentration roughly follows the $\epsilon_T$ dependence and it becomes significantly enhanced for intermediate dopings. In particular, for 
FeNi$_{0.5}$Co$_{0.5}$ system MAE is as large as 180 $\mu$eV/atom. The corresponding anisotropy density constant is equal to 2.4 MJ/m$^3$ which is almost half of the room temperature anisotropy density constant of Nd$_2$Fe$_{14}$B. Moreover, we found that there is a potential for further enhancement of MAE by increasing the tetragonal strain (for example by suitable interstitial doping or epitaxial growth). This is illustrated in Fig. \ref{Fig5} where the MAE of the ordered FeNi$_{0.5}$Co$_{0.5}$ compound is plotted as a function of the $c/a$ ratio (the in-plane lattice parameter was set to the equilibrium value for FeNi$_{0.5}$Co$_{0.5}$). As seen, MAE above 200 $\mu$eV/atom could be realized in this system.

It should be also pointed out that the magnetization of FeNi$_{0.5}$Co$_{0.5}$ is equal to 1.85 T which is significantly larger than that of typical rare-earth-based magnets. Therefore, large energy products can be potentially obtained in FeNi$_{0.5}$Co$_{0.5}$ making this system a very promising material for permanent magnet applications. Experimental efforts to realize FeNi$_{0.5}$Co$_{0.5}$ material by epitaxial growth are currently underway.


In summary, we investigated a new route to introduce and control tetragonality in materials. The idea is based on the Bain transformation and involves combining materials with face- and body-centered symmetries. This concept was illustrated using an example of FeNi$_{1-x}$Co$_x$ alloy obtained by doping the L1$_0$ phase of FeNi with Co. Using first principles electronic structure calculations we demonstrated that the the tetragonal strain increases with doping and it reaches maximum for $x=0.5$. This result was subsequently used to engineer strong MAE in the FeNi$_{1-x}$Co$_x$ alloy. MAE for this system was shown to be a result of complex interplay between tetragonal distortion, electronic concentration and chemical anisotropy mechanisms. We demonstrated that large MAE can be achieved for layered-ordered FeNi$_{1-x}$Co$_x$ alloys. In particular, we identified the FeNi$_{0.5}$Co$_{0.5}$ compound with large MAE of 180 $\mu$eV/atom.

\section*{Acknowledgments}
This work was supported by the Office of Basic Energy Science, Division of Materials Science and Engineering. A. W. acknowledges the support from the Critical Materials Institute, an Energy Innovation Hub funded by the U.S. Department of Energy (DOE). The research was performed at Ames Laboratory, which is operated for the U.S. DOE by Iowa State University under contract \# DE-AC02-07CH11358.


\begin{thebibliography}{99}

\bibitem{Lewis} L. H. Lewis and F. Jim{\'e}nez-Villacorta, Metall. Mater. Trans. A \textbf{44}, 2 (2013).

\bibitem{McCallum} R. W. McCallum, L. H. Lewis, R. Skomski, M. J. Kramer, and I. E. Anderson, Annu. Rev. Mater. Res. \textbf{44}, 451 (2014).

\bibitem{Clarke} R. S. Clarke and E. R. D. Scott, Am. Mineral. \textbf{65}, 624 (1980).

\bibitem{Shima}  T. Shima, M. Okamura, S. Mitani, and K. Takanashi, J. Magn. Magn. Mater. \textbf{310}, 2213 (2007).

\bibitem{Burkert} T. Burkert, O. Eriksson, P. James, S. I. Simak, B. Johansson, and L. Nordstr{\"o}m Phys. Rev. B. \textbf{69}, 104426 (2004).

\bibitem{Burkert2} T. Burkert, L. Nordstr{\"o}m, O. Eriksson, and O. Heinonen Phys. Rev. Lett. \textbf{93}, 027203 (2004).

\bibitem{Andersson} G. Andersson, T. Burkert, P. Warnicke, M. Bj{\"o}rck, B. Sanyal, C. Chacon, C. Zlotea, L. Nordstr{\"o}m, P. Nordblad, and O. Eriksson, Phys. Rev. Lett. \textbf{96}, 037205 (2006).

\bibitem{Yildiz} F. Yildiz, F. Luo, C. Tieg, R. M. Abrudan, X. L. Fu, A. Winkelmann, M. Przybylski, and J. Kirschner, Phys. Rev. Lett. \textbf{100}, 037205 (2008).

\bibitem{Yildiz2} F. Yildiz, M. Przybylski, X.-D. Ma, and J. Kirschner, Phys. Rev. B \textbf{80}, 064415 (2009).

\bibitem{Bain} R. Cuenya, M. Doi, S. Lobus, R. Courths and W. Keune,  Surf. Science  \textbf{493} 338 (2001).

\bibitem{Kotsugi} M. Kotsugi, H. Maruyama, N. Ishimatsu, N. Kawamura, M. Suzuki, M. Mizumaki, K. Osaka, T. Matsumoto, T. Ohkochi, T. Ohtsuki, T. Kojima, M. Mizuguchi, K. Takanashi, and Y. Watanabe, J. Phys.: Condens. Matter \textbf{26}, 064206 (2014).

\bibitem{Blochl} P. E. Bl{\"o}chl, Phys. Rev. B \textbf{50}, 17953 (1994).

\bibitem{Kresse} G. Kresse and J. Hafner, Phys. Rev. B \textbf{48}, 13115 (1993).

\bibitem{Kresse2} G. Kresse and J. Furthm{\"u}ller, Phys. Rev. B \textbf{54}, 11169 (1996).

\bibitem{Antropov} V. P. Antropov, L. Ke, and D. {\AA}berg, Solid State Commun. \textbf{194}, 35 (2014).

\bibitem{FEN} L. Ke, K. D. Belashchenko, M. van Schilfgaarde, T. Kotani, and V. P. Antropov, Phys. Rev. B \textbf{88}, 024404 (2013).




\end{thebibliography}
\end{document}